\documentstyle[11pt,paspconf,epsf]{article}

\def\myfig #1#2#3#4{\par
\epsfxsize=#1 cm
\moveright #2 cm
\vbox{\epsfbox{#3}}
}
\def\et{{et al.}~}
\def\ms{M_{\odot}}
\def\ers{\rm erg~s^{-1}}

\begin{document}

\title{SPECTRAL DEPENDENCE OF THE BROAD EMISSION-LINE REGION IN AGN}

\author{A. Wandel}

\affil{ Racah Institute of Physics, The Hebrew University of Jerusalem 91904, Israel}

\begin{abstract}

We derive a theoretical relation between $R_{BLR}$, the size of the
broad-emission-line region of active galactic nuclei, and the observed soft
X-ray luminosity and spectrum.
We find that in addition to the well known $R_{BLR}\sim L^{1/2}$ scaling,
$R_{BLR}$ depends also on the soft X-ray spectral slope.
Applying this relation
to calculate a predicted BLR radius and emission-line width,
 we show that including the dependence on the spectrum significantly improves
the agreement between the calculated BLR radius and the radius independently
determined from reverberation mapping
 as well as the the agreement between the calculated velocity dispersion
and the observed FWHM of the H$\beta$ line.
We evaluate a theoretical expression for the line width,
providing a physical explanation to the anti-correlation between the
soft X-ray slope and the emission-line width observed in narrow-line
Seyfert galaxies.

\end{abstract}


\section{Introduction}

Recent results from reverberation-mapping of the broad emission-line regions
(BLR) in AGNs indicate that the BLR distance from the central radiation source
roughly scales as $r\propto L^{1/2}$ (Koratkar \& Gaskell 1991; Kaspi et.al.
1996). Our model provides a physical  explanation to this scaling, and predicts
an additional testable dependence on the spectral shape.

Most workers today agree that the line width in AGN is induced by Keplerian
bulk motions. If this is the case, the BLR distance is directly  related to the
line width. A particularly interesting group in this respect are Narrow-Line
Seyfert Galaxies (NLS1), which have narrower emission lines  than ordinary
Seyfert 1 galaxies, and often show steep soft X-ray spectra. In mixed samples
(NLS1 and ordinary AGN) the soft X-ray spectral index is anticorrelated with
the H$\beta$ line width and with the H$\beta$ equivalent width. (Boller, Brandt,
\& Fink 1996; Wang, Brinkmann and Bergeron 1996,  hereafter WBB).

We derive simple analytic relations between soft X-ray continuum spectrum
(luminosity and spectral index) and
the BLR size and the width of the broad emission lines, explaining the
observed anticorrelation between the
 X-ray slope and the line width and equivalent width (Wandel 1997;
Wandel and Boller 1998).
Application of this model to samples of AGN gives a good agreement with
the BLR  sizes determined by
reverberation mapping, and with the observed  FWHM(H$\beta$).

\section{The Model}

 While there is probably a range of physical conditions (ionization
parameter and density etc.) in the broad H$\beta$ line-emitting gas in AGNs,
the emission of each line is dominated by emission in a relatively narrow
optimum range of conditions (Baldwin et al. 1995).  We will therefore
make the assumption that the conditions relevant for H$\beta$ can be approximated
by a single value.
 We show that within the observational constraints
on the far UV and soft X-ray bands of AGN spectra, the  BLR size and the
line width predicted by the model are not very sensitive to the detailed
shape of the spectrum in the EUV band.
We then estimate
the BLR size, and  assuming  Keplerian velocity dispersion we relate it
to the line width. The new element in our scheme is the use of the soft
X-ray spectral index and luminosity to estimate the ionizing spectrum.
 The spectral index enters in the model as follows:
a softer (that is steeper) spectrum has a stronger ionizing power,
and hence the BLR is formed at a larger distance from the central source,
has a smaller velocity dispersion and produces narrower emission lines.
We parameterize the form of the ionizing EUV continuum in terms of the
 the break between the UV and X-ray bands and the measured
soft X-ray slope $\alpha_x$.

\subsection {The BLR radius}

We assume that the spatial extent of the BLR
may be represented by a characteristic size - e.g. the radius at which
the emission peaks.
The physical conditions in the line-emitting gas are largely determined by
the ionization parameter (the ratio of ionizing photon
density to the electron density, $n_e$, e.g. Netzer, 1990)\qquad
$U = Q_{ion} /{4\pi R^2} c n_e  $  
where the ionizing photon flux (number of ionizing photons per unit time) is
$Q_{ion}= \int_{E_0}^\infty F(E) {dE \over E} $,  
$F(E)$ is the luminosity per unit energy,
and $E_0=$1 Rydberg=13.6 eV.
Defining the ionizing luminosity,
$L_{ion} = \int_{E_0}^\infty F(E) dE$
and the mean energy of an ionizing photon,
$\bar E_{ion} \equiv L_{ion}/Q_{ion}$,
the BLR radius may be written as
$$R= \left( {L_{ion}\over 4\pi c\bar E_{ion} U n_e}\right ) ^{1/2}
= 13.4  \left ( {L_{x44}\over U n_{10}  \epsilon_x} \right )^
{1/2} ~~{\rm light-days}\eqno (1)$$
where $n_{10}=n_e/10^{10}~{\rm cm}^{-3}$, $L_{x44}=L_{x}/10^{44}\ers$
is the observed X-ray luminosity,
$ \epsilon =\bar E_{ion}/E_0$ is the mean photon energy in Rydberg,
and $\epsilon_x = \epsilon L_x /L_{ion} = L_x/E_0 Q_{ion}$.
Typical values in the gas emitting the high excitation broad lines are
$U\sim0.1-1$ and
$n_e \sim 10^{10}-10^{11}~{\rm cm}^{-3}$ (e.g. Rees, Netzer and Ferland, 1989),
so that $Un_{10}\sim 0.1-10$.
%


The ionizing flux is dominated by the EUV continuum in the 1-10
Rydberg regime, where most of the ionizing photons are
emitted. Since the the continuum
in this range cannot be observed directly, we try to estimate it by
extrapolation from the nearest observable energy bands.
The far UV spectrum has been observed beyond
the Lyman limit, for about 100 quasars (Zheng et al,
1996), to wavelengths of 600~\AA~ , and for a handful
luminous, high redshift quasars to  wavelengths of
350~\AA.  The average spectrum has a slope of $\alpha\sim 1$ in the
1000--2000~\AA~
band, and below 900~\AA~ it steepens to $\alpha\sim 2$.
We assume that the soft X-ray spectrum can be extrapolated to
lower energies down to some break energy
$E_{b}$, and below the break we take $E^{-2}$.
For the hard X-rays ($E>2$~keV) we use the ``universal'' $E^{-0.7}$ power law,
observed in most AGN with a high energy cutoff at 100~keV.

\subsection {Line width}
Assuming that the line width is induced by  Keplerian motion in the
  gravitational potential of the central mass, the velocity dispersion
corresponding to the
full width at half maximum of the emission lines is given by
%
$v\approx \sqrt {GM/R}$
where $M$ is the mass of the central black hole
and r the distance of the broad line region from the central source.
Equation (1) gives
$$v\approx 1900~ \left ({Un_{10} \epsilon_x\over L_{x44}}\right )^{1/4}
\left ({M\over 10^7 \ms}\right )^{1/2}~{\rm km~s^{-1}}. \eqno (3)$$

In order to relate the unknown mass to the observed
luminosity we assume that
the central mass approximately scales with the luminosity
(Dibai 1981; Joly \et 1985; Wandel \& Yahil 1985; Wandel \& Mushotzky 1986).
In terms of the Eddington ratio these authors find for large AGN samples
$L/L_{Edd}\approx 0.01-0.1$.
Within this distribution, bright objects tend to have slightly
larger $L/M$ ratios than faint ones
(cf. Koratkar and Gaskell 1991)
roughly $L/M\propto L^{1/4}$ or $M\propto L^{3/4}$.
Combining this with the correlation between the optical and the X-ray
luminosities $L_x\propto L_{opt}^{0.75\pm 0.05}$
  (Kriss 1988; Mushotzky and Wandel 1989)
 gives
$$M\approx 7\times 10^7 L_{x44}
\left ({ L/L_{Edd} \over 0.01}\right )^{-1} ~\ms .
\eqno (4)$$

and with eqs. (2) and (3)
the line-width may be related directly to the observed X-ray luminosity
and spectral index:

$$v(FWHM)\approx 5000 ~\eta \epsilon_x ^{1/4} (\alpha_x )
L_{x44}^{1/4} ~~{\rm km~s^{-1}}
 \eqno (5)$$
where
$\eta\equiv (Un_{10})^{1/4}  ( L/L_{Edd} / 0.01 )^{-1/2}$
combines all the unknown parameters;
for the  $Un_{10}\sim 0.1-10$ and  $L/L_{Edd}\sim 0.01-0.1$
stated above we have $\eta \sim 0.2-2$.

\section{Comparison With The Data}
\subsection{BLR radius}

We have Compared the BLR size calculated from the model
with the distance from the central source obtained by reverberation mapping
of the H$\beta$ line,
for a sample of 10 AGN for which reverberation and X-ray data were available
(Figure~2).
As we discuss below, the agreement {\it actually improves} by taking the X-ray
spectral slope into account.

In the model calculations we have used the spectral index
from the power-law fit to the ROSAT 0.1-2.4 keV band and $E_{b}=1$~Ryd,
The horizontal error bars represent the combined error in the
luminosity and in the spectral index.
Allowing for the uncertainty in $Un_{10}\sim 0.1-1$
(represented by the dashed lines in Figure~1), the agreement is
quite good: all the points lie well within these boundaries.
\bigskip

\begin{figure}
\myfig {9.5}{1}{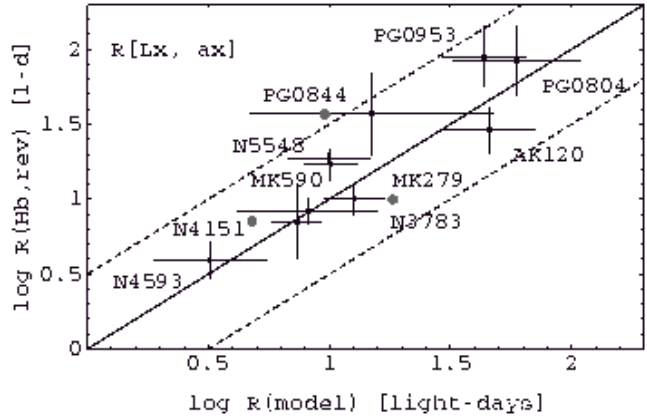}{}
\caption{The BLR size determined from reverberation mapping of the
H$\beta$ line vs. the radius calculated from the model  using
the observed X-ray luminosity and spectral index.
(Wandel 1997).
The grey circles represent the  radius calculated from the model without
taking the spectral dependence into account.
The solid line represents $r_{rev}=r_{model}$ for $Un_{10}=1$
while the dashed parallel lines above and below it
correspond to $Un_{10}=0.1$ and $10$ respectively.
}
\end{figure}
In order to test the significance of our model, we have
 calculated the BLR radii {\it without} taking into account the spectral dependence
(that is, assuming all objects have {\it the same} soft X-ray spectral slope,
which we set to the sample average, $\alpha_x$=1.45).
We find that
the difference between the BLR size (calculated
using only the luminosity scaling) and the reverberation size
is significantly larger than the difference
when the  spectral dependence is taken into account
for three objects
(NGC~4151, PG~0844 and NGC~3783) shown as gray dots in Figure~1.
Those are the objects with the most extreme values of $\alpha_x$.
For the other objects in the sample
the difference is small in both calculations.

\subsection{Line width-spectral index correlation}

Equation (5) predicts an explicit relation between the velocity dispersion
(associated with the line width), the
X-ray luminosity and the spectral index, namely a surface in the
$\alpha_x - v - L_x$ space. For a fixed value of $L_x$ this gives a curve
in the $v- \alpha_x$ plane.
Figure 2 shows such curves of FWHM vs.$\alpha_x$
for several values of the luminosity, $L_{x}=10^{42}-10^{45}\ers$
(Wandel 1997).
Overplotted are the data points -
FWHM(H$\beta$) vs. $\alpha_x$ for a sample of AGN (see below).
The  model seems to reproduce the distribution
of the data very well, and in particular it explains the
observed anticorrelation between
the H$\beta$ line width and the soft X-ray spectral index
and the lack of objects with broad lines and steep soft X-ray spectra.

\begin{figure}
\myfig {9.5}{1}{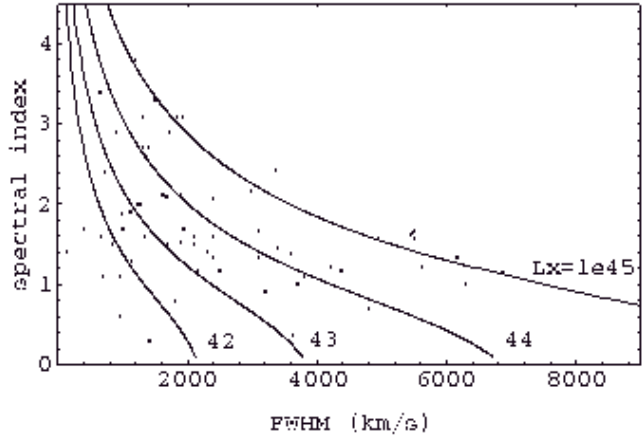}{}
\caption{Theoretical curves of the X-ray spectral index vs. $v$(BLR) for
fixed values of $L_x=10^{42}\ers$ (lower curve) to $L_x=10^{45}\ers$
(upper curve) superposed on the data for the
sample of NLS1s and normal Seyferts. }
\end{figure}

\subsection{Predicted vs. observed line width}

In order to test the model prediction over the three dimensional
$\alpha - v - L_x$ space we compare the observed line width to the
value calculated with equation (5) using the measured X-ray spectral
index and luminosity for a sample consisting of
33 ordinary AGN from Walter \& Fink (1993) combined with 32 NLXGs
(Boller, Brandt, \& ,Fink 1996).

\begin{figure}
\myfig {9.9}{1}{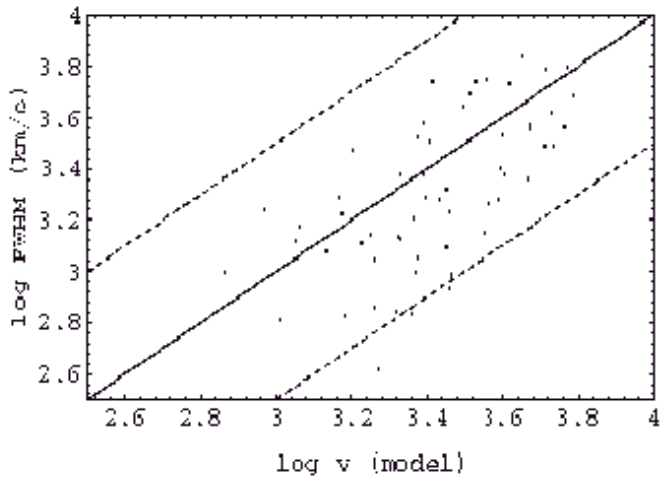}{}
\caption{The observed FWHM(H$\beta$) vs. the v(BLR) calculated from the model
using the observed $L_x$ and spectral index.
The diagonal line represents the equality ($v=FWHM$) for $\eta =0.6$
(see equation [5]),
while the dashed parallel lines above and below the diagonal are for
$\eta=0.2$ and 2 respectively.}
\end{figure}
As can be seen in Figure~3, the agreement is very good, and most of
the objects fall well within the uncertainty strip
of log $FWHM({\rm obs})=\log v({\rm mod}) \pm 0.5$.
When the spectral information is
not taken into account
the correlation is significantly weaker.
The correlation coefficients between the observed and calculated line widths are
$r= 0.533$ and 0.316 respectively.

\section{Equivalent width}
The data show that (WBB)
AGN with a steep soft X-ray spectrum have a lower EW(H$\beta$) than flat-
spectrum AGN.
We now show that also this anti-correlation of the H$\beta$
equivalent width and $\alpha_x$ follows
in the framework of the model.
To see this we recall that
the equivalent width measures the fraction of
the continuum flux reprocessed and emitted in the line.
We have shown that the BLR distance from the central source
increases with the soft X-ray spectral index.
The two-zone photoionization analyses of the BLR (Brotherton \et 1994)
shows that the covering factor in the more extended part BLR part, the
Intermediate Line Region is about 10\% of that in the near BLR, indicating
the integral covering factor decreases with radius.
If such a geometry is common in AGN, a steeper (softer) ionizing spectrum
will lead to a smaller covering factor.
This implies that the equivalent width decreases
with increasing (steepening) spectral index, as observed.
%


\end{document}